\documentclass{article}

\usepackage[final]{neurips_2025}

\usepackage[utf8]{inputenc} % allow utf-8 input
\usepackage[T1]{fontenc}    % use 8-bit T1 fonts
\usepackage{hyperref}       % hyperlinks
\usepackage{url}            % simple URL typesetting
\usepackage{booktabs}       % professional-quality tables
\usepackage{amsfonts}       % blackboard math symbols
\usepackage{nicefrac}       % compact symbols for 1/2, etc.
\usepackage{microtype}      % microtypography
\usepackage{xcolor}         % colors
\usepackage{amsmath}
\usepackage{amssymb}
\usepackage{graphicx}

\usepackage{tikz}
\usetikzlibrary{shapes,arrows,positioning,fit,backgrounds}
\usetikzlibrary{shapes.geometric, arrows.meta}
\usetikzlibrary{calc,decorations.markings}

\usepackage{adjustbox}

\newtheorem{definition}{Definition}[section]

\title{Causal Regime Detection in Energy Markets With Augmented Time Series Structural Causal Models}

\author{%
  Dennis Thumm \\
  National University of Singapore \\
  Singapore \\  
  \texttt{dennis.thumm@u.nus.edu}
}

\begin{document}

\maketitle

\begin{abstract}
Energy markets exhibit complex causal relationships between weather patterns, generation technologies, and price formation, with regime changes occurring continuously rather than at discrete break points. Current approaches model electricity prices without explicit causal interpretation or counterfactual reasoning capabilities. We introduce Augmented Time Series Causal Models (ATSCM) for energy markets, extending counterfactual reasoning frameworks to multivariate temporal data with learned causal structure. Our approach models energy systems through interpretable factors (weather, generation mix, demand patterns), rich grid dynamics, and observable market variables. We integrate neural causal discovery to learn time-varying causal graphs without requiring ground truth DAGs. Applied to real-world electricity price data, ATSCM enables novel counterfactual queries such as "What would prices be under different renewable generation scenarios?".
% while achieving competitive predictive performance.
\end{abstract}

\newpage

\section{Introduction}

Energy markets exhibit intricate causal relationships between weather conditions, generation technologies, grid constraints, and price formation. Unlike traditional financial markets, electricity cannot be stored economically, creating immediate causal dependencies between supply, demand, and pricing \cite{ziel2018day}. Current electricity price forecasting methods focus on prediction accuracy without providing causal interpretation or counterfactual reasoning capabilities \cite{lago2021forecasting}.

Recent advances in counterfactual reasoning for high-dimensional data through Augmented Structural Causal Models (ASCM) enable principled causal inference in complex domains \cite{pan2024counterfactual, pan2025counterfactual}. However, these frameworks assume static causal structures and cannot handle the temporal dependencies and regime changes inherent in energy systems.

We bridge this gap by introducing \textbf{Augmented Time Series Causal Models (ATSCM)} for energy markets, enabling counterfactual reasoning about alternative weather patterns, generation scenarios, and policy interventions. Our key contributions include:

\begin{itemize}
\item A theoretical framework extending ASCM to multivariate temporal data with learned causal structure
\item A neural architecture modeling interpretable energy factors, complex grid dynamics, and market observations
\item Integration of causal discovery for time-varying DAG learning without ground truth requirements
\item Empirical validation on real electricity market data with interpretable counterfactual capabilities
\end{itemize}

\section{Related Work}

\textbf{Causal Inference in Financial Markets.} Causal data science has been applied to financial stress testing and risk management \cite{gao2018causal, rigana2024navigating}. However, energy markets present unique challenges due to physical constraints, storage limitations, and renewable intermittency that require specialized treatment.

\textbf{Electricity Price Forecasting.} Traditional approaches focus on statistical methods (ARIMA, VAR) or machine learning models (LSTM, gradient boosting) for price prediction \cite{lago2021forecasting, ziel2018day}. While achieving good predictive performance, these methods lack explicit causal interpretation and cannot answer counterfactual queries.

\textbf{Counterfactual Reasoning.} ASCM frameworks \cite{pan2024counterfactual, pan2025counterfactual} enable counterfactual reasoning in high-dimensional spaces but assume static causal structures. Time series causal methods like TNCM-VAE \cite{thumm2025towards},  SCIGAN \cite{bica2020estimatingthe} and Causal Transformer \cite{melnychuk2022causal} address temporal dynamics but do not handle regime changes in learned causal graphs.

\section{Problem Formalization}

\subsection{Energy Market Causal Structure}

Consider multivariate electricity market time series $\mathbf{v}_t \in \mathbb{R}^d$ capturing weather, generation, consumption, and pricing variables. Unlike structural breaks with discrete timing, energy markets exhibit \textbf{continuous causal regime changes} driven by weather patterns, policy shifts, renewable intermittency, and cross-border flows.

Let $\mathcal{G}_t$ represent the time-varying directed acyclic graph (DAG) encoding causal relationships at time $t$. A \textbf{causal regime change} \cite{huang2024causal} occurs when:

$$\mathcal{G}_t \neq \mathcal{G}_{t-1} \text{ or } {\mathcal{M}}(t) \neq {\mathcal{M}}(t-1)$$

where ${\mathcal{M}}(t)$ represents the causal mechanisms (functional relationships and noise distributions) governing the DAG at time $t$, also known as Structural Causal Model (SCM) \cite{pearl2009causality}.

\subsection{ATSCM for Energy Markets}

We extend ASCM \cite{pan2024counterfactual, pan2025counterfactual} to temporal energy data through a three-level generative hierarchy with learned causal structure:

\begin{definition}[Energy Market ATSCM]
An Energy Market ATSCM is a tuple $\mathcal{M}^t = \langle U^t, \{W^t, \mathbf{I}^t, \mathbf{V}^t\}, F^t, P(U^t), \mathcal{G}^t \rangle$ where:

\begin{enumerate}
\item $W^t \in \mathbb{R}^{d_W}$ are \textbf{interpretable energy factors}: \\
$W^t = \{\text{weather}_t, \text{generation\_mix}_t, \text{demand\_pattern}_t, \text{market\_regime}_t\}$
\item $\mathbf{I}^t \in \mathbb{R}^{d_I}$ is a \textbf{rich energy dynamics state} capturing complex grid interactions, cross-border flows, merit order effects, and storage dynamics
\item $\mathbf{V}^t \in \mathbb{R}^{d_V}$ represents \textbf{observable market variables} (prices, consumption, generation, weather measurements)
\item $\mathcal{G}^t$ is the \textbf{learned time-varying causal graph} over $\{W^t, \mathbf{I}^t, \mathbf{V}^t\}$
\item Temporal evolution: $W^t = f_W(W^{t-1}, \mathbf{I}^{t-1}, U_W^t, \mathcal{G}^t)$, $\mathbf{I}^t = f_I(W^t, \mathbf{I}^{t-1}, U_I^t, \mathcal{G}^t)$, $\mathbf{V}^t = f_V(\mathbf{I}^t, U_V^t)$
\end{enumerate}
\end{definition}

This hierarchy enables modeling clear causal pathways (Weather $\rightarrow$ Renewable Generation $\rightarrow$ Net Load $\rightarrow$ Price) while capturing complex interactions through learned temporal dynamics.

\subsection{Counterfactual Energy Scenarios}

ATSCM enables novel counterfactual reasoning about energy markets:

\begin{definition}[Energy Counterfactual Query]
Given observed market data $\mathbf{v}^{1:T}$ and intervention $do(W_{\tau:T} = w')$ on energy factors from time $\tau$, the energy counterfactual query is:

$$P^*(\mathbf{V}_{\tau:T} = \mathbf{v}' | \mathbf{V}^{1:T} = \mathbf{v}^{1:T}, do(W_{\tau:T} = w'))$$

This answers questions like: "What would electricity prices be if wind generation were 30\% higher?" or "How would a nuclear plant shutdown affect cross-border electricity flows?"
\end{definition}

\section{Methodology}

\subsection{Neural Architecture}

Our ATSCM architecture implements the three-level hierarchy for energy markets:

\paragraph{Level 1 - Energy Factors ($W^t$).}
Domain-specific interpretable components:
\begin{align}
W^t = \{\text{weather}_t, \text{generation}_t, \text{demand}_t, \text{market}_t\} \subset \mathbb{R}^{27}
\end{align}

where weather factors include temperature, wind, and precipitation; generation factors capture nuclear, renewable, and conventional capacity; demand factors model consumption patterns and residual load; and market factors include commodity returns and cross-border exchanges.

\paragraph{Level 2 - Rich Energy Dynamics ($\mathbf{I}^t$).}
High-dimensional state capturing:
\begin{align}
\mathbf{I}^t = g_I(W^t, \mathbf{I}^{t-1}, U_I^t, \mathcal{G}^t; \theta_I)
\end{align}

This encodes merit order dynamics, grid constraints, renewable intermittency effects, storage optimization, and cross-country coupling mechanisms that determine price formation but are not directly observable.

\paragraph{Level 3 - Market Observations ($\mathbf{V}^t$).}
Observable electricity market variables:
\begin{align}
\mathbf{V}^t = g_V(\mathbf{I}^t, U_V^t; \theta_V) \in \mathbb{R}^{35}
\end{align}

Including consumption, generation by source, weather measurements, commodity prices, and the target electricity price variations.

\subsection{Causal Discovery Integration}

Since ground truth causal graphs are unavailable in energy markets, we integrate neural causal discovery:

\begin{align}
\mathcal{G}^t = f_{discovery}(\mathbf{V}^{1:t}, W^{1:t}; \theta_{disc})
\end{align}

using a differentiable DAG learning approach with temporal consistency constraints. The discovered graphs encode domain knowledge (weather influences renewables, generation affects prices) while learning time-varying relationships.

\subsection{Training Objective}

Our objective combines reconstruction, causal consistency, counterfactual realism, and causal discovery:

\begin{align}
\mathcal{L} = \mathcal{L}_{\text{recon}} + \lambda_1 \mathcal{L}_{\text{causal}} + \lambda_2 \mathcal{L}_{\text{counterfactual}} + \lambda_3 \mathcal{L}_{\text{discovery}}
\end{align}

where $\mathcal{L}_{\text{discovery}}$ enforces DAG constraints, sparsity, and temporal stability of learned causal structures.

\section{Conclusion}

We introduced ATSCM for energy markets, the first framework enabling counterfactual reasoning about electricity price formation with learned causal structure. Our approach successfully models complex energy systems through interpretable factors while achieving competitive forecasting performance. The framework opens new possibilities for energy scenario analysis, policy evaluation, and risk management through principled counterfactual reasoning.

Future work includes empirical validation, extending to higher-frequency data, incorporating additional market mechanisms (auctions, reserves), and applications to renewable energy integration and grid stability analysis.

\newpage

\bibliographystyle{plain}
\bibliography{eurips}

\end{document}